\documentclass[aps,prl,showpacs,twocolumn,superscriptaddress]{revtex4}

\usepackage{graphicx}
\usepackage{amsmath}
\usepackage{amssymb}
\usepackage{dcolumn}% Align table columns on decimal point
\usepackage{bm}% bold math

\begin{document}

\title{Two qubit conditional quantum logic operation in a single self-assembled quantum dot}

\author{S.~J.~Boyle}
\email{s.boyle@shef.ac.uk}
\affiliation{Department of Physics and Astronomy, University of
Sheffield, Sheffield, S3 7RH, United Kingdom}

\author{A.~J.~Ramsay}
\email{a.j.ramsay@shef.ac.uk}
 \affiliation{Department of Physics
and Astronomy, University of Sheffield, Sheffield, S3 7RH, United
Kingdom}

\author{F.~Bello}
\affiliation{Department of Physics
and Astronomy, University of Sheffield, Sheffield, S3 7RH, United
Kingdom}

\author{H.~Y.~Liu}
\affiliation{Department of Electronic and Electrical Engineering,
University of Sheffield, Sheffield, S1 3JD, United Kingdom}

\author{M.~Hopkinson}
\affiliation{Department of Electronic and Electrical Engineering,
University of Sheffield, Sheffield, S1 3JD, United Kingdom}

\author{A.~M.~Fox}
\affiliation{Department of Physics and Astronomy, University of
Sheffield, Sheffield, S3 7RH, United Kingdom}

\author{M.~S.~Skolnick}
\affiliation{Department of Physics and Astronomy, University of
Sheffield, Sheffield, S3 7RH, United Kingdom}

\date{\today}% It is always \today, today,
             %  but any date may be explicitly specified

\begin{abstract}
The four-level exciton/biexciton system of a single semiconductor
quantum dot acts as a two qubit register. We experimentally
demonstrate an exciton-biexciton Rabi rotation conditional on the
initial exciton spin in a single InGaAs/GaAs dot. This forms the
basis of an optically gated two-qubit controlled-rotation (CROT)
quantum logic operation where an arbitrary exciton spin is
selected as the target qubit using the polarization of the control
laser.
\end{abstract}

% PACS, the Physics and Astronomy
                             % Classification Scheme.
%\keywords{Suggested keywords}%Use showkeys class option if keyword
                              %display desired
\maketitle

Two qubit gates that operate on a `target' qubit conditional on
the state of the `control' qubit, such as the controlled rotation
(CROT) gate, provide the basic tool for entangling and
disentangling qubits, and are therefore a critical component of
any quantum  processor \cite{Barenco_prl}. One possible
realization of a quantum processor is based on the use of
semiconductor quantum dots, where the qubit is encoded in the
presence or absence of an exciton, and manipulated using
picosecond laser pulses. To date, considerable experimental
progress has been made in the coherent optical control of a single
exciton qubit
\cite{Bonadeo_sci,Zrenner_nat,Stievater_prl,Stufler_prb}, and a
two qubit gate has been reported for two excitonic qubits hosted
in a single GaAs interface dot \cite{Li_sci}. In the latter case,
a CROT-gate is realized by driving a Rabi rotation on the
exciton-biexciton transition, similar to proposals in ref.
\cite{Biolatti_prl}.

In this letter, we report on the conditional coherent optical
control of two exciton qubits hosted in a single InGaAs/GaAs
quantum dot. This is achieved by observing an exciton-biexciton
Rabi rotation conditional on the initial state of the exciton
spin, where the $\pi$-pulse acts as a controlled-rotation (CROT)
gate. In contrast to ref. \cite{Li_sci}, we investigate the case
of a CROT-gate where the polarization of the control laser is used
to address both exciton-biexciton transitions simultaneously. In
similar atomic systems arranged in a 3-level
$\Lambda$-configuration, when two lasers are used to address each
transition, arbitrary superpositions of the lower energy states
can be coupled to the upper state \cite{Kral_rmp}. In a quantum
dot context, this provides a potential tool for selecting
arbitrary exciton spin superpositions as the target and control
qubits, providing a new degree of control over the two-qubit
operation.

\begin{figure}
\begin{center}
\includegraphics[width=0.4\textwidth]{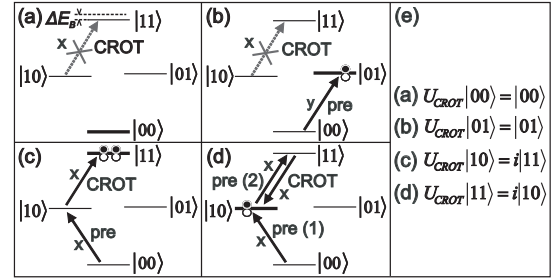}
\end{center}
\caption{
Implementation of a CROT gate on two exciton-based qubits.
The QD ground state $\vert 00 \rangle$, the two orthogonally
 polarized exciton $(X^{0})$ states $\vert 01 \rangle$, $\vert 10 \rangle$,
 the biexciton ($2X^{0}$) state $\vert 11 \rangle$,
 and the biexciton binding energy $\Delta E_{B}$ are
 indicated. The CROT operation is performed by an
 x-polarized $\pi$-pulse resonant with the $X^{0}-2X^{0}$
 transition. Cases (a)-(d) show the CROT operation
 on all four pure qubit states. %(a) The QD is in the
 %ground state, so the CROT pulse is off-resonant
 %and is not absorbed. (b) A y-polarized preparation
 %pulse prepares the $\vert 01 \rangle$ state; the
 %CROT pulse is not absorbed due to polarization selection
 % rules. (c) An x-polarized preparation pulse prepares
 % the $\vert 10 \rangle$ state, so the CROT pulse
 % excites the $\vert 10 \rangle\leftrightarrow\vert 11 \rangle$
  %transition. (d) The $\vert 11 \rangle$ state is
  %prepared and the CROT pulse returns the QD to the
  %$\vert 10 \rangle$ state.
  (e) The CROT truth table.
}
\label{fig_1}
\end{figure}

The experiments are performed on a single QD embedded in an
n-i-Schottky diode structure, at a temperature of $\sim
10~\mathrm{K}$, full details of which can be found in
ref.~\cite{Kolodka_prb}. The dot is excited by up to two
spectrally narrow laser pulses, with a Gaussian pulse shape and
FWHM of 0.2~meV. A photocurrent detection method
\cite{Zrenner_nat} is used, where under the action of an applied
electric-field, the electron-hole pairs tunnel from the dot
resulting in a change in the photocurrent.
%Fixed-mask pulse shaping techniques, described in ref.~\cite{weiner2000},

The QD exciton-biexciton system may be considered as two
excitonic-qubits labeled by their spin-states. There are 4 states:
the ground state $\vert 00 \rangle$, two orthogonally polarized
single exciton states $\vert 01 \rangle$ and $\vert 10 \rangle$,
and the biexciton state $\vert 11 \rangle$, as illustrated in fig.
~\ref{fig_1}. The electron-hole exchange interaction causes the
energy eigenstates of the exciton to be linearly polarized along
the crystal lattice axes, designated the x and y axes. This
results in strong selection rules for linear polarization and a
fine structure splitting of $ 10-100~\mathrm{\mu eV}$. Due to the
Coulomb interaction, the ground state to exciton ($0-X^{0}$) and
exciton to biexciton ($X^{0}-2X^{0}$) transitions are separated by
a biexciton binding energy $\hbar\delta_{B}$, of a few meV. %These
%effects enable the selective excitation of four separate
%transitions using spectrally narrow polarized pulses: $\vert 00
%\rangle \leftrightarrow \vert 01 \rangle$, $\vert 00 \rangle
%\leftrightarrow \vert 10 \rangle$, $\vert 01 \rangle
%\leftrightarrow \vert 11 \rangle$ and
% $\vert 10 \rangle \leftrightarrow \vert 11 \rangle$.
The CROT operation flips the `target' qubit if the `control' qubit
is in the $\vert 1\rangle$ state, according to the truth table
shown in fig. \ref{fig_1}(e). The gate can be realized by an
x-polarized
 pulse with a pulse-area of $\pi$, resonant with
 the $X^{0}-2X^{0}$ transition, labelled the CROT
 pulse. The effect of this pulse on all four
 pure qubit states is shown in fig. \ref{fig_1}(a)-(d).
  (a) The dot is initially in the
 ground state $\vert 00 \rangle$. Hence the CROT pulse
 is not resonant with any available transition and
 is not absorbed. (b) A y-polarized
 preparation pulse (pre-pulse) with a pulse area of $\pi$,
  resonant with the $0-X^{0}$ transition, precedes the
  CROT pulse, preparing the y-polarized exciton state
  $\vert 01 \rangle$. The CROT pulse is not absorbed
  due to polarization selection rules. (c)An x-polarized
  pre-pulse prepares the x-polarized
  exciton state $\vert 10 \rangle$, and the CROT pulse
  excites the $\vert 10 \rangle \leftrightarrow \vert 11 \rangle$
  transition. (d) Having prepared the $\vert 11\rangle$ state as in
  (c) using the first half of a $2\pi$-pulse, the second half of
  the pulse drives the system back to $\vert 10\rangle$.

\begin{figure}
\begin{center}
\includegraphics[width=0.4\textwidth]{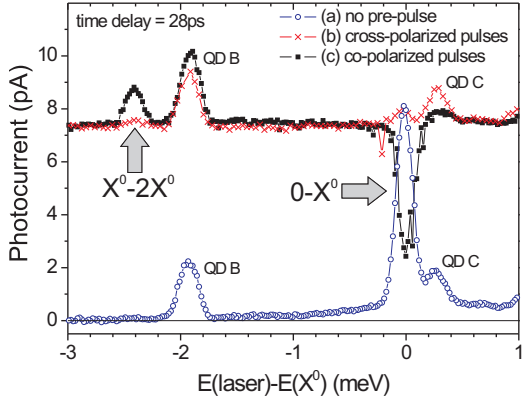}
\end{center}
\caption{ Photocurrent spectra. $(\circ)$ The single-pulse
measurement shows the ground state to exciton transition
$(0-X^{0})$ and two features from neighboring dots (QD B and QD
C). For two-pulse measurements, a linearly polarized preparation
pulse prepares the exciton state. $(\bullet)$ With co-polarized
pulses, the exciton-biexciton transition $(X^{0}-2X^{0})$ and
quenching of the exciton are observed. $(\times)$ With
cross-polarized pulses the $(X^{0}-2X^{0})$ transition is
suppressed by polarization selection rules. } \label{fig_spec}
\end{figure}

Single pulse and two-pulse measurements are made on a single QD
with a $0-X^{0}$ transition energy of 1.302 eV, at a reverse bias
of 0.6 V. In all measurements, `pre-pulse' refers to a $\pi$-pulse
resonant with the $0-X^{0}$ transition. The spectra in
fig.~\ref{fig_spec} show photocurrent as a function of the
detuning with respect to the $0-X^{0}$ transition of a
$\pi$-pulse. (a) The single pulse measurement shows the $0-X^{0}$
transition. Two weak features, labelled B and C, are also observed
due to the excitation of other nearby QDs. For two-color
measurements, this pulse is preceded by the pre-pulse, which
prepares an exciton state. Both pulses are linearly polarized
along the crystal axes. (b) In the case of co-polarized pulses, an
additional peak corresponding to the $X^{0}-2X^{0}$ transition is
observed at a detuning of $\hbar\delta_{B}=-2.41~\mathrm{meV}$
\cite{det_eff}. At zero detuning a dip is observed, as the two
pulses act as a $2\pi$ Rabi rotation, first creating an
 exciton and then returning the dot to the ground state.
  (c) For cross-polarized pulses, polarization selection
   rules suppress the $X^{0}-2X^{0}$ transition.

\begin{figure}
\begin{center}
\includegraphics[width=0.4\textwidth]{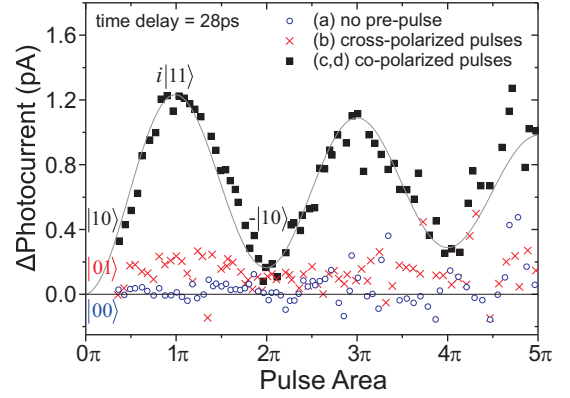}
\end{center}
\caption{ Conditional exciton-biexciton ($X^{0}-2X^{0}$) Rabi
rotation (Change in photocurrent vs pulse area for a pulse
resonant with the $X^{0}-2X^{0}$ transition). $(\circ)$ The single
pulse measurement shows no oscillation, as the pulse is
off-resonant, as in fig.~\ref{fig_1}(a). For two-pulse
measurements, a linearly polarized preparation pulse prepares the
$X^{0}$ state. $(\bullet)$ With co-polarized pulses, a Rabi
rotation between the $X^{0}$ and $2X^{0}$ states is observed,
corresponding to fig.~\ref{fig_1}(c) and (d) for pulse areas of
$\pi$ and $2\pi$ respectively. $(\times)$ Polarization selection
rules suppress the Rabi rotation for cross-polarized pulses, as in
fig.~\ref{fig_1}(b). The corresponding logic states are indicated.
} \label{fig_rabi}
\end{figure}

To demonstrate the truth table of the CROT operation, we measure
the $X^{0}-2X^{0}$ Rabi rotation for all four pure qubit input
states. Conditional Rabi rotation measurements are shown in
fig.~\ref{fig_rabi} as the change in photocurrent as a function of
the pulse area of the CROT pulse, which is resonant with the
$X^{0}-2X^{0}$ transition and is polarized along one of the
crystal axes. A background photocurrent, linear in power, is
subtracted from all data. (a) The measurement without a pre-pulse
shows a flat line, because the input state is the QD ground state
$\vert 00 \rangle$ and the CROT pulse is off-resonant, as in
fig.~\ref{fig_1}(a). For two-color measurements, a pre-pulse, also
polarized along one of the crystal axes, precedes the CROT pulse,
preparing an exciton state. (b) In the measurement with
cross-polarized pulses, an exciton is prepared polarized
orthogonally to the control pulse, yielding an input state of
$\vert 01 \rangle$. A flat response is measured, as polarization
selection rules suppress the $X^{0}-2X^{0}$ transition as in
fig.~\ref{fig_1}(b). (c,d) With co-polarized pulses, an exciton is
prepared with the same polarization as the control pulse. The
input state is then $\vert 10 \rangle$, and more than two periods
of a weakly damped Rabi rotation between the $\vert 10 \rangle$
and $\vert 11 \rangle$ states are observed. This corresponds to
fig.~\ref{fig_1}(c) and \ref{fig_1}(d) for pulse areas of $\pi$
and $2\pi$ respectively.
%An estimate of the relative dipole moments, $\mu_{X},\mu_{2X}$, of
%the $0-X^{0}$ and $X^{0}-2X^{0}$ transitions, is made from
%comparisons of the Rabi rotations. A value of
%$\frac{\mu_{2X}}{\mu_{X}}=0.95\pm0.03$ is found, comparable to the
%value of $\frac{\mu_{2X}}{\mu_{X}}=0.85\pm0.08$ reported for GaAs
%QDs \cite{Stievater_prb}.
These experiments demonstrate the truth
table of the CROT gate for pure qubit input states.

In the long term, high-fidelity gates will be needed to realize
fault-tolerant quantum information processing. It is therefore
useful to evaluate the performance of this CROT gate in an InAs
dot, and compare it to that with a GaAs dot \cite{Li_sci}. Since
the potential coherence times of InAs dots are far superior to
GaAs dots (600~ps \cite{Borri_prl} versus 50~ps \cite{Li_sci}),
much higher gate fidelities are expected. One feasible comparison
is to calculate the fidelity using experimentally determined
parameters in a 4-level atom model as described in the
supplementary information of ref. \cite{Li_sci}. For the data
presented in fig.~\ref{fig_rabi}, the exciton coherence time is
limited by electron tunnelling to $T_2=130\pm 20~\mathrm{ps}$,
giving a fidelity of $ 0.87\pm 0.04$, compared to $0.7$ for  GaAs
interface dots where $T_2$ is limited by the radiative lifetime to
$46~\mathrm{ps}$ \cite{Li_sci}. If a similar calculation is made
using state-of-the-art values for photocurrent detection with InAs
dots (laser bandwidth 0.4~meV, $T_2= 320~\mathrm{ps}$)
\cite{Stufler_prb} a fidelity of $0.97$ is in prospect. We note
that in the region of interest (pulse-area $\Theta<2\pi$) the
model is in good agreement with the Rabi rotation data, where the
signal at $2\pi$ returns to its value at $0\pi$ to within 15\%.
However, the model does not account for intensity damping effects
\cite{Wang_prb}, and should therefore be treated cautiously.
Overall, the comparison supports the expectation that much higher
fidelity quantum gates are possible with InAs dots compared with
GaAs dots.

%In ref. \cite{Li_sci}, Li {\it et~al} suggest that a higher
%fidelity CROT operation can be achieved using an InAs quantum dot,
%due to their longer coherence times compared with GaAs dots. It is
%not currently feasible to measure the gate fidelity, and a
%calculation based on a 4-level atom with experimentally determined
%decay rates would be misleading, since recent observations of
%decreased intensity damping for Rabi rotations driven with longer
%laser pulses contradicts this model \cite{Wang_prb}. To aid a
%direct experimental comparison of the CROT gates for GaAs and InAs
%dots we define the ``return-to-zero''  as
%$RTZ=\frac{I(2\pi)}{I(\pi)}$, where $RTZ_{GaAs}\sim 0.3\sim
%1-F_{GaAs}$ \cite{Li_sci} versus $RTZ_{InAs}=0.15\pm 0.03$. This
%comparison strongly suggests that InAs QDs have more favorable
%fidelities than GaAs QDs. Here the coherence time is limited by
%electron tunnelling to  $T_1= 130\pm 20 ~\mathrm{ps}$, longer
%times $T_2=320~\mathrm{ps}$ have been observed in photocurrent
%measurements \cite{Stufler_prb}, suggesting that InAs dots can
%achieve far lower RTZ values than the one reported here.

 The polarization of the control
laser may be used to drive both exciton-biexciton transitions
simultaneously, and by analogy with similar atomic
$\Lambda$-transitions \cite{Kral_rmp}, should provide an
additional degree of control over the two-qubit operation. We thus
present a study of the polarization properties of the
exciton-biexciton Rabi rotation. We consider the case of two laser
pulses, designated (1) and (2), with carrier frequencies
on-resonance with the ($0-X^0$) and ($X^0-2X^0$) transitions
respectively. Both pulses are spectrally narrow so that they
excite only one set of transitions. The control Hamiltonian in the
rotating frame of the states $\vert
00\rangle,\vert\uparrow\rangle,\vert\downarrow\rangle,\vert
11\rangle$, where the spin up/down exciton states
$\uparrow$/$\downarrow$ are used for ease of presentation, is
given by:

\begin{equation*}
\hat{H}= \frac{1}{2} \left( \begin{array}{cccc}
0 & \Omega_{+}^{(1)}  & \Omega_{-}^{(1)} & 0 \\
\Omega^{(1)\ast}_{+} & 0 & \delta_{\text{fs}} & \Omega_{-}^{(2)} \\
\Omega^{(1)\ast}_{-} & \delta_{\text{fs}} & 0 & \Omega_{+}^{(2)} \\
0 & \Omega^{(2)\ast}_{-} & \Omega^{(2)\ast}_{+} & 0
\end{array} \right)
\end{equation*}

\noindent where $\hbar=1$,  and $\delta_{\text{fs}}$ is the
fine-structure splitting, and $\Omega_{\pm}^{(\alpha)}$ are the
$\sigma_{\pm}$ circularly polarized components of the complex Rabi
frequency of each laser. For the sake of clarity, we neglect
coherence loss. If the pulses have no spectral or temporal
overlap, and are much shorter than the fine-structure period, the
control Hamiltonian may be interpreted as a time-sequence of
rotations $\hat{U}_{\gamma}$.

For the pre-pulse the control Hamiltonian reduces to:

\begin{equation*}
\hat{H}_{1}= \frac{1}{2}\vert 00\rangle[ \Omega_{+}^{(1)}\langle
\uparrow\vert +\Omega_-^{(1)}\langle \downarrow\vert] + h.c.\equiv
\frac{\Omega_1}{2}\vert 00\rangle \langle A_1\vert + h.c.
\end{equation*}

\noindent where $\Omega_{\alpha}=\sqrt{\vert
\Omega_+^{\alpha}\vert^2+\vert\Omega_-^{\alpha}\vert^2}$ is the
effective Rabi frequency of pulse ($\alpha$). Here the
vacuum-exciton transitions form a 3-level V-configuration
\cite{Wang_prl}. The pre-pulse drives a Rabi rotation between the
vacuum state $\vert 00\rangle$ and the exciton-spin superposition
state $\vert A_1\rangle$ labelled the ``active'' state, leaving
the orthogonal ``inactive'' state $\vert \bar{A}_1\rangle$
untouched. Hence the full polarization of a pre-pulse with
pulse-area of $\pi$ is imprinted on the exciton spin.

During the inter-pulse time interval the fine-structure drives a
rotation $\hat{U}_{\mathrm{fs}}$ between exciton spin up and down
states \cite{Tartakovskii_prl}. For the CROT pulse the control
Hamiltonian reduces to:

\begin{equation*}
\hat{H}_{2}= \frac{1}{2}[ \Omega_{-}^{(2)}\vert \uparrow\rangle
+\Omega_+^{(2)}\vert \downarrow\rangle]\langle 11\vert +
h.c.\equiv \frac{\Omega_2}{2}\vert A_2\rangle\langle 11\vert +
h.c.
\end{equation*}

\noindent Here the exciton-biexciton transitions form a 3-level
$\Lambda$-configuration, where the CROT pulse drives a Rabi
rotation between the ``active''  exciton spin superposition state
$\vert A_2 \rangle$, and the biexciton state $\vert 11\rangle$,
leaving the ``inactive'' $\vert \bar{A}_2\rangle$ state untouched.
In essence the polarization can select any arbitrary exciton spin
as the `target' $\vert \bar{A}_2\rangle$ and `control' $\vert
A_2\rangle$ qubits, and this can be used as a control tool. For
example, the polarization of a CROT pulse with a pulse-area of
$\pi$ can select an exciton spin superposition to project into the
biexciton state, providing a means of detecting the exciton spin.
Alternatively, full optical control of the exciton spin can be
achieved as follows. A CROT-pulse with a pulse-area of $2\pi$
imparts a detuning-dependent phase-shift of up to $\pi$
\cite{Ramsay_arxiv,Economu_prl} on $\vert A_2\rangle$ with respect
to $\vert \bar{A}_2\rangle$, and since any exciton spin
superposition may be selected as $\vert A_2\rangle$, full optical
control of the exciton spin can be achieved using the polarization
of the pulse \cite{Wang_prl}. This results in a gate time limited
by the duration of the control pulse, far faster than two-pulse
techniques where the exciton spin is controlled using the
fine-structure beat \cite{Bonadeo_sci,Stufler_prb}.

\begin{figure}
\begin{center}
\includegraphics[width=0.48\textwidth]{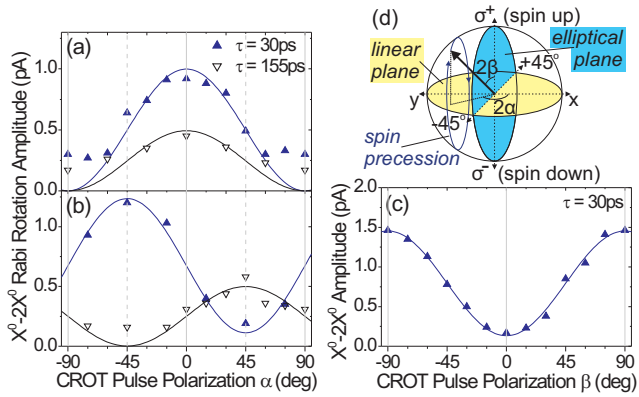}
\end{center}
\caption{ %Polarization dependent Rabi rotation measurements. %A
%pre-pulse, a $\pi$-pulse resonant with the ground state to exciton
%($0-X^{0}$) transition, is followed by the CROT pulse resonant
%with the exciton-biexciton ($X^{0}-2X^{0}$) transition.
%The amplitude of the $X^{0}-2X^{0}$ Rabi rotation is measured as a
%function of the CROT pulse polarization. This gives a measurement
%of the exciton spin component with which the CROT pulse couples.
%(A),(B) For linearly polarized pulses, $\alpha$ gives the
%orientation of polarization with respect to the x axis. For a
%short time delay between the pulses, $\tau$=30ps, the Rabi
%rotation amplitude has a sinusoidal dependency on $\alpha_{pre}$
%and $\alpha_{control}$. In (A) ($\alpha_{pre}=0^{\circ}$) this
%dependency is preserved at longer delay, as a stable spin state is
%prepared. In (B) ($\alpha_{pre}=-45^{\circ}$), the relationship is
%inverted at a time delay of $\tau$=155ps, corresponding to a
%$\sim1\pi$ fine structure oscillation. (C) The measurement is
%performed with an elliptically polarized control pulse
%$(\Omega^{-},\Omega^{+}) \equiv \Omega [\cos (\beta), \sin
%(\beta)]$ and a $\sigma^{+}$-polarized pre pulse. A dependency on
%$\beta_{control}$ is observed in agreement with the model (see
%main text). (D) Bloch sphere representation of exciton spin, with
%spin precession due to fine structure.
Amplitude of the $X^{0}-2X^{0}$ Rabi rotation measured as a
function of the polarization of the CROT pulse. (a,b)Linearly
polarized CROT-pulse probes exciton created with (a) x, (b)
$-45^{\circ}$ linearly polarized pre-pulse. (c) An elliptically
polarized CROT-pulse probes exciton created by $\sigma_+$
polarized pre-pulse. Full lines (a)-(c), are fits to data using:
$\Delta PC\propto\vert \langle A_2 \vert\hat{U}_{\mathrm{fs}}\vert
A_1\rangle\vert^2$. (d) Poincar\'{e} sphere representation of
exciton spin, with spin precession due to fine structure. }
\label{fig_pol}
\end{figure}

To test this model, we study the dependence of the $0-X^0$, and
$X^0-2X^0$ Rabi rotations on the polarizations of the  pre and
CROT pulses. Figure~\ref{fig_pol}(d) shows a Poincar\'{e} sphere
representation of the polarization where
$(\Omega_+,\Omega_-)=\Omega(
e^{i\alpha}\cos{\beta},e^{-i\alpha}\sin{\beta})$. The period of
the $0-X^0$ and $X^0-2X^0$ Rabi rotations are independent of
polarization. For time-resolved measurements (not shown), where
both the pre and CROT pulses are $\pi$-pulses, fine-structure
beats with a period of 320 ps are observed. %The fine-structure splitting increases with
%vertically applied electric-field from
%($\hbar\delta_{\mathrm{fs}}=\frac{h}{320\pm
%10~\mathrm{ps}}\rightarrow\frac{h}{230\pm 10~\mathrm{ps}}$ , for
%$V_{rev}=0.6\rightarrow 0.8\mathrm{V}$, or $F=57.8\rightarrow 66.4
%~\mathrm{kV.cm^{-1}}$).

To test that the CROT-pulse drives a Rabi rotation between the
``active'' exciton spin superposition $\vert A_2\rangle$ and the
biexciton $\vert 11\rangle$ states, we measure the amplitude of
the Rabi rotation as a function of the polarization of the CROT
pulse, in both the linear and elliptical planes of the
Poincar\'{e} sphere (fig.~\ref{fig_pol}d), and compare with the
model: $\Delta PC\propto\vert \langle A_2
\vert\hat{U}_{\mathrm{fs}}\vert A_1\rangle\vert^2$.

({\it Linear plane}) The pre-pulse is linearly polarized along the
x-axis, to excite an energy eigen-state, and the Rabi rotations
are measured as a function of linear polarization $\alpha_{crot}$
with ($\alpha_{pre}=0, \beta_{pre}=\beta_{crot}=+45^{\circ}$).
Figure ~\ref{fig_pol}(a) presents the results. A cosine dependence
is observed where maximum signal occurs when the pulses are
co-linearly polarized. This is true for all time-delays since
there is no precession of the exciton spin.  Next, for the results
presented in fig.~\ref{fig_pol}(b), a $-45^{\circ}$ linearly
polarized pre-pulse is used, and the amplitude recorded as a
function of the CROT linear polarization
($\alpha_{pre}=-45^{\circ},
\beta_{pre}=\beta_{crot}=+45^{\circ}$). At short time-delay
($\delta_{\mathrm{fs}}\tau<\pi/2$) maximum signal again occurs for
co-linearly polarized excitation. However,  at
$\tau=155~\mathrm{ps}$ the maximum now occurs for cross-linear
polarization due to the fine-structure rotation. In
fig.~\ref{fig_pol}(a-c) the solid-lines show the good fits to data
using $\Delta PC\propto\vert \langle A_2
\vert\hat{U}_{\mathrm{fs}}\vert A_1\rangle\vert^2$ , where the
amplitudes of the oscillations are the only fitting parameters.

({\it Elliptical plane}) To probe the elliptically polarized
plane, a circularly polarized pre-pulse is used
$(\alpha_{pre}=\alpha_{crot}=45^{\circ}, \beta_{pre}=0)$, and the
amplitude of the Rabi rotation measured as a function of the
ellipticity angle of the CROT-pulse $\beta_{crot}$. At short
time-delay, the maximum is observed for cross-circular excitation,
see fig.\ref{fig_pol}(c). The measurements show close agreement
with the model for both planes of the Poincar\'{e} sphere,
implying that the full polarization of the pre-pulse is stored in
the exciton spin, and then a selected component is projected into
the biexciton state by the CROT-pulse. This further implies that
the polarization of the CROT pulse selects an exciton spin
superposition to couple optically to the biexciton state.

In conclusion, we have demonstrated a CROT quantum logic gate for
two excitonic qubits in a single InGaAs/GaAs dot, with a high
fidelity of $0.87\pm 0.04$. Furthermore, we find that the
polarization of the control pulse may be used to select arbitrary
exciton spin superpositions as the target and control qubits. This
property may be used to construct an operator which combines
elements of the CROT and bit-swap operations into a single control
pulse, offering a more efficient control sequence than a series of
CROT and single qubit operations.

This work was funded by EPSRC UK GR/S76076 and the QIPIRC UK.

\bibliographystyle{apsrev}
\bibliography{BoyleCROT}

\begin{thebibliography}{1}
\bibitem{det_eff}
The photocurrent detection efficiency  is limited
by the hole tunneling rate
\cite{Kolodka_prb}. Hence the $X^0-2X^0$ signal, which involves tunnelling of two holes,
 is weaker than $0-X^0$ at this bias.
\end{thebibliography}

\end{document}